# Observation of two-photon interference effect with single non-photon-number resolving detector

Heonoh Kim,[1] Sang Min Lee,[2] Osung Kwon,[3] and Han Seb Moon[1,*]

[1]*Department of Physics, Pusan National University, Geumjeong-Gu, Busan 46241, South Korea*
[2]*Korea Research Institute of Standards and Science, Daejeon 34113, South Korea*
[3]*National Security Research Institute, Daejeon 34044, South Korea*
*\*Corresponding author: hsmoon@pusan.ac.kr*

**Multiphoton interference effects can be measured with a single detector when two input photons are temporally well separated when compared with the dead time of the single-photon avalanche detector. Here we experimentally demonstrate that the Hong-Ou-Mandel interference effect can be observed with a single non-photon-number resolving detector via a time-delayed coincidence measurement of successive electrical signals from the detector. The two-photon interference experiment is performed by utilizing temporally well-separated pairwise weak coherent pulses and the interference fringes are successfully measured with a high visibility in the range of the limited upper bound for the weak coherent photon source.**

The Hong-Ou-Mandel (HOM) interference effect is a well-known two-photon quantum interference phenomenon, which has been considered as two-photon coalescence to the same output spatial mode of a beamsplitter (BS) when two identical photons are simultaneously incident on the BS-input ports [1]. This characteristic feature has played an important role in the experimental quantum optics in studying fundamental quantum mechanics as well as in the field of quantum information science and technology [2-7]. For the observation of the HOM interference fringe, a coincidence measurement with two single-photon detectors positioned at the BS-output ports is usually employed because of the impossibility of distinguishing between a single photon and two photons impinging the detector. Few experimental demonstrations of the direct observation of the HOM effect with a single detector have been performed by utilizing the photon-number resolving detector [8] and the single-photon-sensitive intensified sCMOS camera system [9]. In addition, two-photon coalescence effects have also been observed with the aid of two-photon counting statistics of the single-photon detection rates that rely on the nonlinear response of a single-photon avalanche photodiode [10,11]. However, the two-photon interference effect can be measured with a single non-photon-number resolving (NPNR) detector when temporally well-separated pairwise pulsed photons are employed as the input photon source.

In this Letter, we report on a novel experimental demonstration of the observation of the HOM-type interference effect in the output signal of only a single NPNR detector via time-delayed coincidence measurement of two successive electrical signals from the detector. For this demonstration, we employed temporally well-separated weak coherent pulses (WCPs) as the input photon source to perform the HOM-type two-photon interference experiment, which is implemented in a polarization-based Michelson interferometer. As a result, the two-photon interference effect is successfully measured with a high visibility in the range of the limited upper bound for the WCP photon source. In particular, the two-photon interference effect and the HOM-type dip fringes as shown in the coincidence measurement using two single-photon detectors are simultaneously observed with single-detector output signals.

Figure 1 shows the schematic of a method to measure the two-photon interference effect with only one single-photon avalanche detector via coincidence measurement of successive electrical signals from the single detector. Here, if we consider time-delayed coincidence counts of two detectors or a single detector, as shown in Fig. 1, the most dominant terms of the input state can be expressed in the temporally well-separated pairwise two-photon states as the form

$$|\Psi\rangle = \frac{1}{2}\left[\hat{a}_1^\dagger \hat{a}_1^\dagger(\Delta t) + \hat{a}_1^\dagger \hat{a}_2^\dagger(\Delta t) + \hat{a}_2^\dagger \hat{a}_1^\dagger(\Delta t) + \hat{a}_2^\dagger \hat{a}_2^\dagger(\Delta t)\right]|0,0\rangle, \quad (1)$$

where $a_i^\dagger$ denotes the photon creation operator and $\Delta t$ is the separation time between two consecutive pulses, and the subscripts represent the two spatial modes of the BS-input ports. Eq. (1) describes the pairwise two-photon state in the case where one photon is in a pulse of the spatial mode 1 (2) and the other one is in the pulse followed by $\Delta t$ of the spatial mode 2 (1) or 1 (2). For the HOM-type experiment, input photons are prepared in a superposed state with temporally well-separated pairwise WCPs (dashed arrow lines in the BS-input stage). When this type of pairwise WCPs is incident on the balanced BS, the two input photons in each pulse exit the same output port of the BS regardless of the separation time between two pulses, as shown in Fig. 1(a), which behavior is similar to that observed when two identical photons are simultaneously incident on the BS [1,12,13]. Here, if we ignore the case of two sequential pulses in the same spatial mode with one photon per pulse at the BS-input port, i.e. pairwise path-entangled states consisting of the two-photon amplitudes, $a_1^\dagger a_1^\dagger(\Delta t)$ and $a_2^\dagger a_2^\dagger(\Delta t)$, the HOM-type interference effect is observed at the BS-output port with perfect visibility [13].

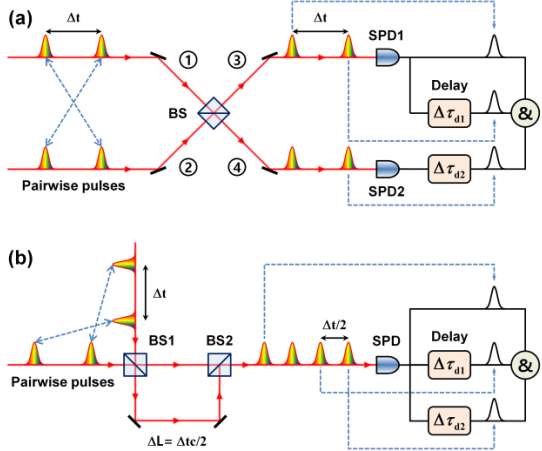

Fig. 1. Observation of a two-photon interference effect via coincidence measurement of successive signals from a single detector. (a) Scheme to observe interference effect with one and two detectors simultaneously. (b) Scheme to observe interference effect with only one detector via mixing sequential pairwise pulses. Photon source, WCPs with center wavelength of 775 nm and repetition rate of 20 MHz; $\Delta t$, pulse period; BS, 50/50 beamsplitter; SPD, single-photon avalanche detector; $\Delta\tau_{d1}$ and $\Delta\tau_{d2}$, electrical delay; &, coincidence electronics.

As a result, two two-photon amplitudes, $a_3^\dagger a_3^\dagger(\Delta t)$ or $a_4^\dagger a_4^\dagger(\Delta t)$, are observed at the single output port via time-delayed coincidence measurement of the consecutive electrical signals from the single detector SPD1 with accompanying electrical delay $\Delta\tau_{d1}$ as well as the coincidences at SPD1 and SPD2 with $\Delta\tau_{d2}$ ($\Delta\tau_{d1} = \Delta\tau_{d2} = n\Delta t$, where $n$ denotes an integer). Although the NPNR detector output cannot directly reveal the two-photon interference event, it is possible to observe the interference fringe when two input pulses are well-separated from each other when compared with the dead time of the single-photon detector. Next, we consider the second setup, shown in Fig. 1(b), in which an optical delay of $\Delta L = \Delta t c / 2$, where $c$ denotes the speed of light in vacuum, is employed to mix the two output modes of BS1 with a temporal delay of a half-pulse duration. In this case, the HOM-type coincidence peak and dip fringes are simultaneously observed with an electrical delay time of $\Delta\tau_{d1} = n\Delta t$ ($n$ denotes an integer) and $\Delta\tau_{d2} = m\Delta t / 2$ ($m > 1$ denotes an odd integer).

Figure 2 shows the real experimental setup to measure the HOM-type two-photon interference effect with employing temporally well-separated pairwise WCPs. Our setup is roughly similar to those of previous studies using thermal light in continuous or pulsed modes [14,15] and ultrashort WCPs [12,16]. In this Letter, we used a WCP train from a mode-locked picosecond fiber laser as the input photon source with a 3.5-ps pulse duration at a 775-nm center wavelength with a 20-MHz repetition rate. In the setup, the laser output pulses are highly attenuated by several linear polarizers, and then coupled into a single-mode fiber. The polarizing beamsplitter (PBS) is used to define the polarization direction for the polarization-based Michelson interferometer, and the superposed input state consisting of the temporally well-separated pairwise WCPs is prepared via a half-wave plate with its axis oriented at 22.5° followed by the PBS. Two quarter-wave plates are placed in the two interferometer arms to rotate the polarization direction with its axis oriented at 45°. Thus the two spatial modes (1 and 2) in Eq. (1) are demonstrated as two polarization modes (H: horizontal, V: vertical) in this experiment. Two kinds of pairwise two-photon states can be generated within the two interferometer arms: one has the two amplitudes $|1\rangle_H |1(\Delta t)\rangle_V$ and $|1\rangle_V |1(\Delta t)\rangle_H$, the other has $|1\rangle_H |1(\Delta t)\rangle_H$ and $|1\rangle_V |1(\Delta t)\rangle_V$ [13]. The former is used in our experiment to observe the HOM-type effect; on the other hand, the latter has to be removed to obtain a high-visibility interference fringe. However, this kind of the path-entangled pairwise two-photon states cannot be removed in the case where WCPs are employed because of the statistical property in the photon-number distribution [14-19]. Alternatively, the path-entangled pairwise state can be disabled to have no influence on the observation of the phase-insensitive HOM-type interference fringe via phase randomization between the two interferometer arms [12-14]. Thus, the maximum attainable fringe visibility is restricted less than 0.5 for the WCPs, which is caused by the noninterfering constant coincidences from the phase-randomized path-entangled two-photon states.

To measure the HOM-type interference effect with temporally separated WCPs, the path-length difference between the two interferometer arms is adjusted by moving the mirror M2, which is mounted on the motorized translation stage, while one of the mirrors M1 is affixed to the PZT actuator in order to randomize the relative phase between the two paths. The phase randomization is necessarily required to negate the phase-sensitive interference effect of the path-entangled pairwise two-photon state as well as the interference fringe of the one-photon state [12-14]. In our experiment, two kinds of measurement setups are employed to measure the two-photon interference effect. The first is a combination of a BS and two linear polarizers P1 and P2, as shown in Fig. 2(b), which is a typical setup to analyze the polarization correlations by post-selective measurement behind the BS for orthogonally polarized input photons. The two single-photon detectors (D1 and D2) record the pairwise sequential WCPs by utilizing an electrical delay $\Delta\tau_{d2}$ of 50 ns. At the same time, the single detector D1 measure the two-photon interference peak fringe via time-delayed coincidence measurement of consecutive electrical signals from the detector with time delay $\Delta\tau_{d1}$ of 50 ns, which is longer than the dead time of the detector. For the second setup of Fig. 2(c), an optical delay line is employed to introduce a 25-ns temporal interval by use of a 5-m-long single-mode fiber, where the vertically polarized components of the temporally separated input photons are round tripped around the PBS and then mixed again in the middle between the horizontal components by a 25-ns time delay. Once the two-photon interference occurs at the PBS, the two spatially bunched and temporally well-separated photons exit either the transmission port with horizontal polarization or the reflection port with vertical polarization. As a result, the two-photon interference effect (coincidence "peak" fringe) is observed in the time-delayed coincidences at D and D ($n\Delta t$) while the HOM-type coincidence "dip" fringe is observed at D and D ($m\Delta t/2$). In this measurement, the electrical delay times of $\Delta\tau_{d1}$ and $\Delta\tau_{d2}$ are set at 100 ns and 125 ns, respectively.

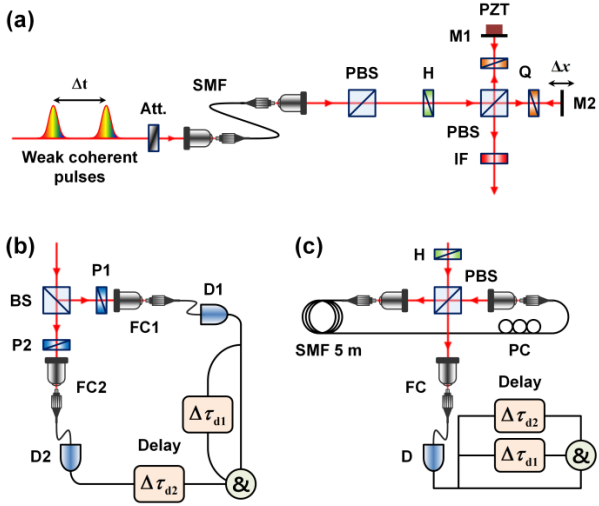

Fig. 2. Experimental setup to measure the two-photon interference effect with single detector employing temporally well-separated WCPs. $\Delta t$, pulse period of 50 ns, Att, attenuator; SMF, single-mode fiber; PBS, polarizing beamsplitter, H, half-wave plate; Q, quarter-wave plate; M1 and M2, mirrors; PZT, piezo-electric transducer; IF, interference filter with a 1-nm bandwidth; BS, 50/50 beamsplitter; P1 and P2, linear polarizers; FC, single-mode fiber couplers; PC, polarization controller; D, single-photon detector (SPCM-AQR-4C, Perkin Elmer); $\Delta\tau_{d1}$ and $\Delta\tau_{d2}$, electrical delay lines; &, coincidence electronics. Optical path-length difference is introduced by varying $\Delta x$ of the M2 position. See text for further experimental details.

The two-photon interference fringes are observed as a function of the path-length difference $\Delta x$, as shown in Fig. 3. In the experiment, the measured single and coincidence counting rates were approximately 300 kHz and 4.5 kHz, respectively. In the first measurement setup, shown in Fig. 2(b), the coincidence dip (peak) fringe was observed at the two detectors D1 and D2 ($\Delta\tau_{d2}$) with the combination of polarization analyzer angles at $\theta_{P1}=+45°$ and $\theta_{P2}=+45°$ ($\theta_{P2}=-45°$), as shown in Figs. 3(a) and 3(b). On the other hand, only the peak fringes were observed at the single detector D1 and D1 ($\Delta\tau_{d1}$) via coincidence measurement of the two consecutive electrical signals from the detector. Here, the coincidence peak measured at the two detectors does not accompany the dip in the single detector for the polarization post-selective measurement behind the BS for orthogonally polarized input photons. The filled squares and circles represent the experimental data and the solid lines are the theoretical fits. The error bars represent one standard deviation of the measured coincidence counting rates, which is nearly the same as the symbol size due to the higher counting rate in our experiment. The measured fringe visibilities were found to be 0.49 ± 0.01 and 0.47 ± 0.01, respectively, for Fig. 3(a), and 0.47 ± 0.01 and 0.47 ± 0.01, respectively, for Fig. 3(b). The Gaussian-shaped fringe width was approximately 0.6 mm, which value is related to the spectral bandwidth of the interference filter used in our experiment.

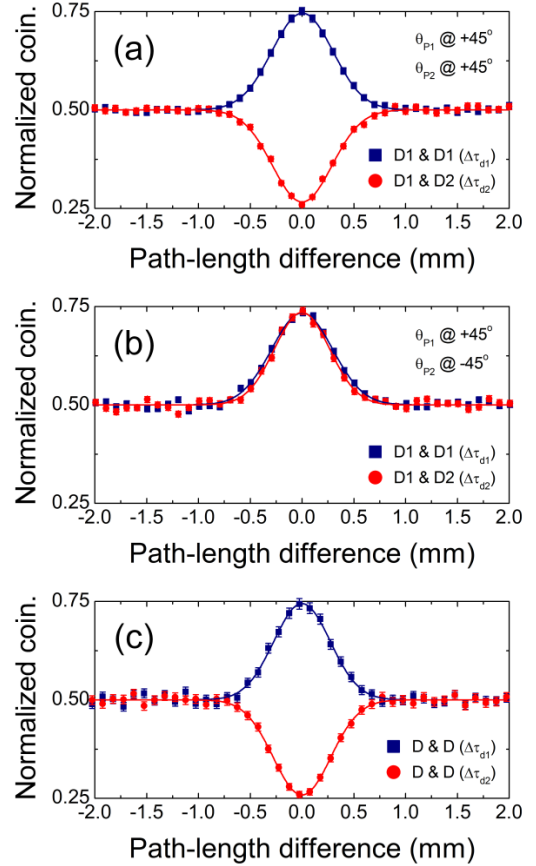

Fig. 3. Experimental results showing normalized coincidences as a function of the path-length difference of $\Delta x$. The HOM-type interference fringes are measured with one and two detectors for polarization analyzer angle combinations of (a) $\theta_{P1}/\theta_{P2}=+45°/+45°$ and (b) $\theta_{P1}/\theta_{P2}=+45°/-45°$, with accompanying electrical delays of $\Delta\tau_{d1}$ and $\Delta\tau_{d2}$. (c) The coincidence peak/dip fringes are simultaneously measured with only one detector. Error bars represent the square root of the measured coincidence counting rates.

An interesting result is observed with the second measurement setup, shown in Fig. 2(c), in which the coincidence peak and dip fringes are simultaneously measured with only one detector, as shown in Fig. 3(c). The two-photon interference effect is observed in the coincidence peak between two successive detector-output signals with a time delay of 100 ns. The HOM-type dip fringe is observed when the coincidence electronics record two consecutive signals from the detector with a time delay of 125 ns which corresponds to the two spatially separated pairwise pulses at the PBS-output port. The measured visibilities were found to be 0.49 ± 0.01 for the peak and dip fringes. In our experiment, the maximally attainable rate is limited to 20 MHz due to the rather long dead time of the single-photon detector. If a single-photon detector with a very short dead time is utilized, our measurement-device-efficient detection method can be very useful for application in quantum information technology together with the resource efficient source of entangled photons [20]. Here, we remark that the dead-time issue of the single-photon detector can be overcome with current technology [21,22]. The issue of the unbalanced overall efficiencies of the single-photon detection systems can also be overcome by using only one detector as shown in this Letter.

In conclusion, we have experimentally demonstrated the observation of two-photon interference fringes with a single NPNR detector. We performed HOM-type interference experiments using temporally well-separated pairwise WCPs in a polarization-based Michelson interferometer. Simultaneous measurements of the two-photon coincidence peak and dip fringes were carried out by two time-delayed coincidence measurements of the successive electrical signals from the single detector. Although our experiment provides a proof-of-principle demonstration of the possibility of the observation of the two-photon interference fringes using only one non-photon-number resolving single-photon detector with a fairly long dead time, multiphoton interference experiments and high-speed detection can also be implemented with current detector technology.

**Funding.** Measurement Research Center (MRC) Program; National Research Foundation of Korea (NRF) (2015R1A2A1A05001819, 2016R1D1A1B03936222, 2014R1A1A2054719)